%% file: 19-asilomar-v6.tex
\documentclass[10pt,conference,twoside]{IEEEtran}

\usepackage{cite}
\usepackage[T1]{fontenc}
\usepackage{graphicx}
\usepackage{amssymb}
\usepackage{amsmath}
\usepackage{amsthm}

\usepackage{microtype}
\usepackage{hyperref}
\usepackage{url}
\usepackage{balance}
\usepackage{xcolor}
\usepackage{subfigure}

\input{vmr-symbols-vecbold}
\input{standard-macros}

\input{defs}

\allowdisplaybreaks 

\IEEEoverridecommandlockouts

\begin{document}


\title{Design Trade-offs for Decentralized Baseband Processing in Massive MU-MIMO Systems}
\author{\IEEEauthorblockN{Kaipeng Li$^\text{1}$, James McNaney$^\text{1}$, Chance Tarver$^\text{1}$, Oscar Casta\~neda$^\text{2}$,\\ Charles Jeon$^\text{2}$, Joseph R. Cavallaro$^\text{1}$, and Christoph Studer$^\text{2}$} \\
\IEEEauthorblockA{\em $^\text{1}$Department of Electrical and Computer Engineering, Rice University, Houston, TX}
\IEEEauthorblockA{\em $^\text{2}$Department of Electrical and Computer Engineering, Cornell University, Ithaca, NY} 
\thanks{The work was supported in part by Xilinx, Inc. and by the US NSF under grants ECCS-1408370, CNS-1717218, CNS-1827940, ECCS-1408006, CCF1535897, CCF-1652065, and CNS-1717559.}
}

\maketitle


\fussy

\begin{abstract}
Massive multi-user (MU) multiple-input multiple-output (MIMO) provides high spectral efficiency by means of spatial multiplexing and fine-grained beamforming. However, conventional base-station (BS) architectures for systems with hundreds of antennas that rely on centralized baseband processing inevitably suffer from (i) excessive interconnect data rates between radio-frequency circuitry and processing fabrics, and (ii) prohibitive complexity at the centralized baseband processor. Recently,  decentralized baseband processing (DBP) architectures and algorithms have been proposed, which mitigate the interconnect bandwidth and complexity bottlenecks.
This paper systematically explores the design trade-offs between error-rate performance, computational complexity, and data transfer latency of DBP architectures under different system configurations and channel conditions. 
Considering architecture, algorithm, and numerical precision aspects, we provide practical guidelines to select the DBP architecture and algorithm that are able to realize the full benefits of massive MU-MIMO in the uplink and downlink.
\end{abstract}


\section{Introduction}
Massive multi-user (MU) multiple-input multiple-output (MIMO) will be a key technology component in fifth-generation (5G) and future wireless communication systems~\cite{5gbe}.
The idea of this technology is to equip the infrastructure base-stations (BSs) with hundreds of antenna elements while serving tens of user equipments (UEs) simultaneously and in the same frequency band.
The presence of a large number of antennas at the BSs enables fine-grained beamforming, which provides higher spectral efficiency than traditional, small-scale MIMO systems~\cite{JH_13, mimo_overview}. 
However, na\"ively scaling up small-scale MIMO systems to large antenna arrays will  inevitably result in a range of practical implementation challenges~\cite{Li_JETCAS17} which must be resolved before deploying massive MU-MIMO in practice.

\subsection{Challenges with Centralized Baseband Processing}
The excessively large amount of raw baseband data that must be transferred between the BS antenna array and the baseband processing backhaul is among the most critical challenges that arise with large antenna arrays~\cite{puglielli2015scalable,Li_JETCAS17, quantizeMM, eric_dp2018}.
As an example, a $256$-BS-antenna massive MU-MIMO system with $12$-bit digital-to-analog converters (DACs) supporting a bandwidth of $80$\,MHz requires raw baseband data rates from and to the radio-frequency (RF) chains that approach $1$\,Tb/s. Such high data rates not only exceed the limits of existing interconnect technology by large margins, e.g., that of the  Common Public Radio Interface (CPRI)~\cite{cpri}, but also push chip input/output (I/O) interfaces, power dissipation, and processing capabilities of modern computing fabrics, such as graphics processing units (GPUs), field-programable gate arrays (FPGAs), or application-specific integrated circuits (ASICs), to their limits. 

\begin{figure}[tp]
\centering
\includegraphics[width=0.95\columnwidth]{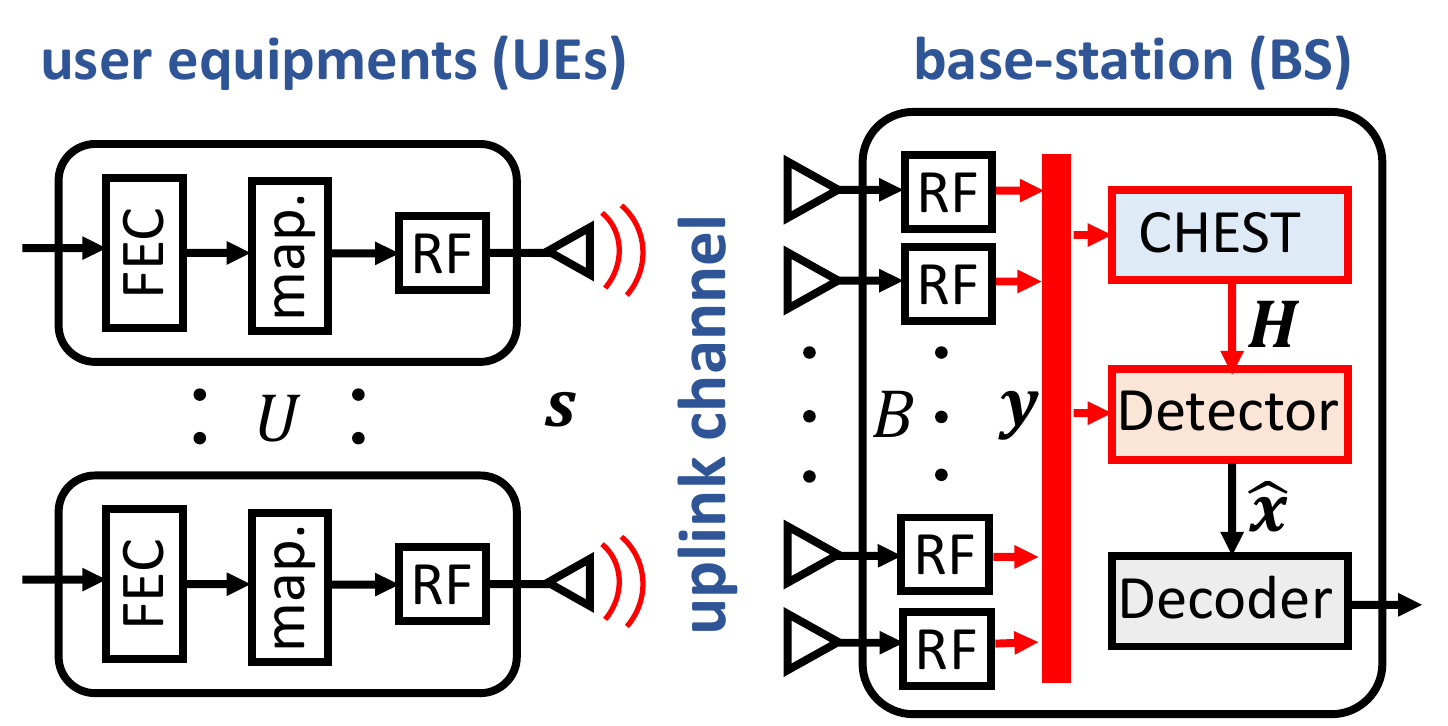}
\caption{Illustration of a base-station (BS) in the massive MU-MIMO uplink with a centralized baseband processing architecture.
The large amount of raw baseband data that must be 
transferred between the RF chains and the computing fabric (illustrated with red color)
is a key implementation challenge. 
} 
\label{fig:cbp}
\end{figure}

Figure~\ref{fig:cbp} illustrates the interconnect bandwidth and complexity bottlenecks (highlighted with red color) in the massive MU-MIMO uplink (UEs transmit to BS) with conventional \emph{centralized baseband processing} architectures. Note that we face the same bottlenecks in the downlink (BS transmits to UEs).
Although one could resort to maximum ratio combining (MRC) for fully distributed data detection and maximum ratio transmission (MRT) for fully distributed precoding, MRC and MRT result in low spectral-efficiency compared to that of more complex centralized algorithms, such as minimum mean-square error (MMSE) equalization~\cite{JH_13} or linear Wiener filter precoding~\cite{asilomar18}.
Existing massive MU-MIMO testbeds, such as the BigStation~\cite{bigstation} and the Lund testbed~\cite{lund}, incorporate centralized MMSE and ZF algorithms while parallelizing the computation workload across subcarriers. While this approach mitigates the complexity bottleneck, it still suffers from the interconnect and chip I/O bottlenecks as one must transfer raw baseband data from and to all antennas.

\subsection{Decentralized Baseband Processing}

To effectively avoid the bottlenecks of centralized baseband processing and to enable scalability to massive MU-MIMO architectures with hundreds or even thousands of antenna elements, recent work
~\cite{Li_JETCAS17, eric_dp2018, lund_dp2019} introduced \emph{decentralized baseband processing} (DBP). 
DBP partitions the antenna array into smaller clusters, each associated with separate RF circuitry and baseband processing fabrics. Each antenna cluster only connects with the associated computing fabrics which perform local baseband processing tasks, such as channel estimation, data detection in the uplink, and precoding in the downlink. 
While consensus-sharing based methods have been proposed for these tasks~\cite{Li_JETCAS17, Li_Asilomar_16, Li_GlobalSIP_16}, the iterative exchange of information among clusters suffers from data transfer latency which negatively affects the design efficiency.
Recently, the references~\cite{isit17,tsp19,Li_Asilomar_17} and~\cite{asilomar18} proposed \emph{feedforward DBP architectures} for uplink detection and downlink precoding, respectively. Such feedforward architectures avoid the repeated exchange of information among clusters, which mitigates data transfer latency issues~\cite{asilomar18}. Furthermore, the theoretical analysis in~\cite{tsp19} revealed that feedforward equalization architectures with linear algorithms are able to achieve the same or similar spectral efficiency as their centralized counterparts. 

\subsection{Contributions}
While the literature describes a number of \emph{feedforward} DBP architectures and algorithms~\cite{tsp19,asilomar18,eric_dp2018}, 
a systematic trade-off analysis under different system configurations and channel conditions is missing. Such a trade-off analysis, however, is critical to making design decisions for practical massive MU-MIMO systems that rely on DBP.
This paper focuses on such trade-offs at different levels for {feedforward} DBP architectures and algorithms which avoid iterative consensus exchange. 
In \fref{sec:tradearch}, we analyze the performance and data transfer bandwidth trade-offs of two feedforward architectures dependent on system configuration and channel conditions, and we show that the channel coherence time is a critical design factor for \emph{architecture-level} trade-offs.
In \fref{sec:tradealg}, we investigate the performance and complexity trade-offs of different feedforward equalization and precoding algorithms, and we show that channel coherence time and channel reciprocity are important design factors for \emph{algorithm-level} trade-offs. 
In \fref{sec:tradeprec}, we study the performance and efficiency trade-offs when reducing the arithmetic precision of data transfers and numerical computations.
In  \fref{sec:conclusions}, we conclude the paper and summarize practical design guidelines.

\section{Architecture Trade-offs}
\label{sec:tradearch}
We now study the performance and data transfer bandwidth trade-offs  in the uplink and downlink for DBP feedforward architectures dependent on the  channel's coherence time. 

\subsection{System Models and Architectures}
\label{sec:sysmodel}
We now detail the uplink and downlink channel models and provide details on DBP feedforward architectures.

\subsubsection{Uplink}
\label{sec:ulmodel}
The uplink consists of $U$ single-antenna UEs transmitting data to a BS with $B$ antennas. 
The data of the~$U$ UEs is contained in the transmit vector $\bmx^\text{ul}\in\setO^U$, where $\setO$ is the constellation (e.g., 16-QAM). The BS receives the vector $\mathbf{y}^\text{ul}\in\mathbb{C}^B$, which we model using the baseband input-output relation $\bmy^\text{ul}=\bH^\text{ul}\bmx^\text{ul}+\bmn^\text{ul}$, where $\bH^\text{ul}\in\mathbb{C}^{B\times U}$ represents the channel matrix, and $\bmn^\text{ul}\in \mathbb{C}^{B}$ is i.i.d.\ complex circularly-symmetric Gaussian noise with variance $\No$ per entry. 

\begin{figure}[t]
\centering
\subfigure[Partially decentralized (PD)]{\includegraphics[width=0.24\textwidth]{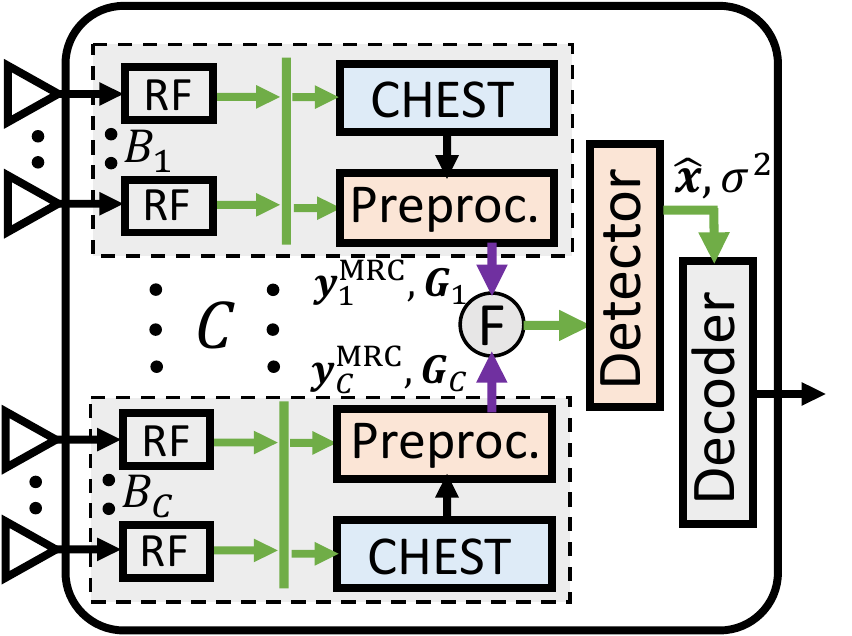}\label{fig:2a}}
\subfigure[Fully decentralized (FD)]{\includegraphics[width=0.227\textwidth]{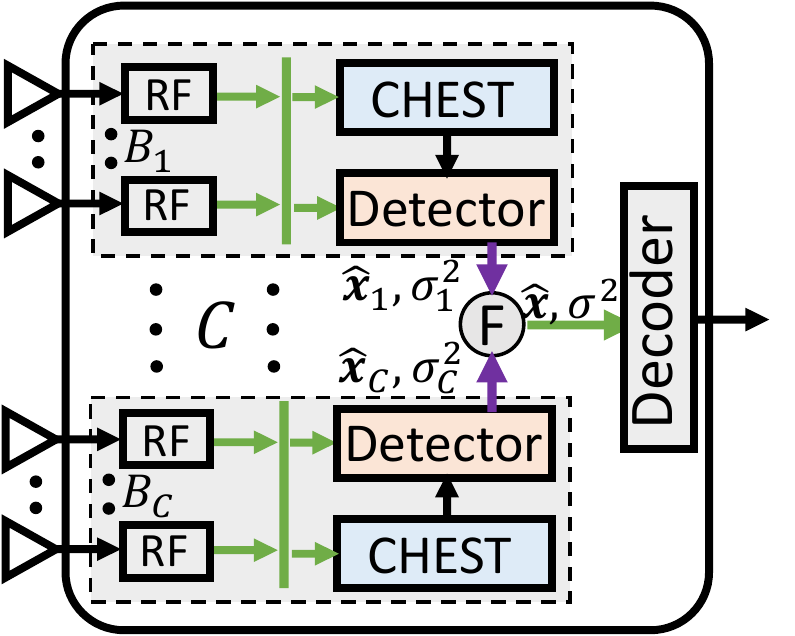}\label{fig:2b}}

\caption{Decentralized feedforward architectures for the massive MU-MIMO uplink. (a) The 
partially-decentralized (PD) architecture separates preprocessing and data detection. (b) The fully-decentralized (FD) architecture performs data detection at each antenna cluster followed by centralized data fusion.}
\label{fig:fdarch}
\end{figure}

The feedforward DBP architectures proposed in~\cite{isit17,tsp19} partition the $B$ BS-antennas into $C$ antenna clusters; Fig.~\ref{fig:fdarch} illustrates two DBP architectures. 
The $c^\text{th}$ cluster has $B_c$ antennas where $B=\sum^C_{c=1}{B_c}$, and each cluster receives its own $\bmy_c^\text{ul}\in\mathbb{C}^{B_c}$. The  input-output relation at each cluster is then
\begin{align}
\bmy_c^\text{ul}=\bH_c^\text{ul}\bmx^\text{ul}+\bmn_c^\text{ul}, \quad c=1,2,\ldots,C.
\end{align}
Here, $\bH_c^\text{ul}\in \mathbb{C}^{B_c\times U}$ is the local channel matrix, a sub-matrix of $\bH^\text{ul}$, and $\bmn_c^\text{ul}\in \mathbb{C}^{B_c}$ represents the local noise vector at cluster $c$. We focus on two distinct feedforward DBP architectures put forward in~\cite{isit17}: the partially-decentralized (PD) architecture and fully-decentralized (FD) architecture shown in~\fref{fig:2a} and~\fref{fig:2b}, respectively. 
Both architectures allow each BS cluster to estimate the local matrix $\bH_c^\text{ul}$ and perform local preprocessing based on $\bH_c^\text{ul}$ and the local receive vector $\bmy_c^\text{ul}$ to form the local Gram matrix $\bG_c^\text{ul}={\bH_c^\text{ul}}^H\bH_c^\text{ul}$ and matched filter output $\bmy_c^\text{MRC}={\bH_c^\text{ul}}^H\bmy_c$. 
The PD architecture performs fusion of $\bG_c^\text{ul}$ and $\bmy_c^{\text{MRC}}$ at the centralized processing unit to produce the equalization output $\hat{\bmx}^\text{ul}$. The FD architecture performs local equalization and requires data fusion of the local equalization outputs $\hat{\bmx}_c^\text{ul}$ and noise variance $\hat{\sigma}_c^\text{ul}$ to produce the equalization output $\hat{\bmx}^\text{ul}$. See~\cite{isit17,tsp19} for the details.

\subsubsection{Downlink System}
\label{sec:dlmodel}
In the downlink, the BS computes the precoding vector $\bmx^\text{dl}=\bP^\text{dl}\bms^\text{dl}$, where $\bmx^\text{dl}\in\mathbb{C}^B$, $\bP^\text{dl}\in\complexset^{B\times U}
$ is the precoding matrix and $\bms^\text{dl}\in\setO^U$ is the transmit data vector. At the UEs, the receive vector $\bmy^\text{dl}\in\mathbb{C}^U$ is given by $\bmy^\text{dl}=\bH^\text{dl}\bmx^\text{dl}+\bmn^\text{dl}$, where $\bH^\text{dl}\in\mathbb{C}^{U\times B}$ and $\bmn^\text{dl}\in \mathbb{C}^{U}$ are the downlink channel matrix and noise vector, respectively. 

Analogous to the uplink, reference \cite{asilomar18} proposed feedforward DPB architectures for the downlink, where each of the $C$ antenna clusters uses the local downlink channel matrix $\bH_c^\text{dl}={\bH_c^\text{ul}}^T$, the transpose of $\bH_c^\text{ul}$, assuming channel reciprocity, in order to form the local beamforming vector $\bmx_c^\text{dl}$. See \cite{asilomar18} for the details. The input-output relation at each cluster is
\begin{align}
\bmy_c^\text{dl}=\bH_c^\text{dl}\bmx^\text{dl}+\bmn_c^\text{dl}, \quad \bH_c^\text{dl}\in \mathbb{C}^{U\times B_c}, \quad c=1,2,\ldots,C.
\end{align}

\begin{figure}[t]
\centering
\subfigure[BER: PD-MMSE vs. FD-MMSE]{\includegraphics[width=0.24\textwidth]{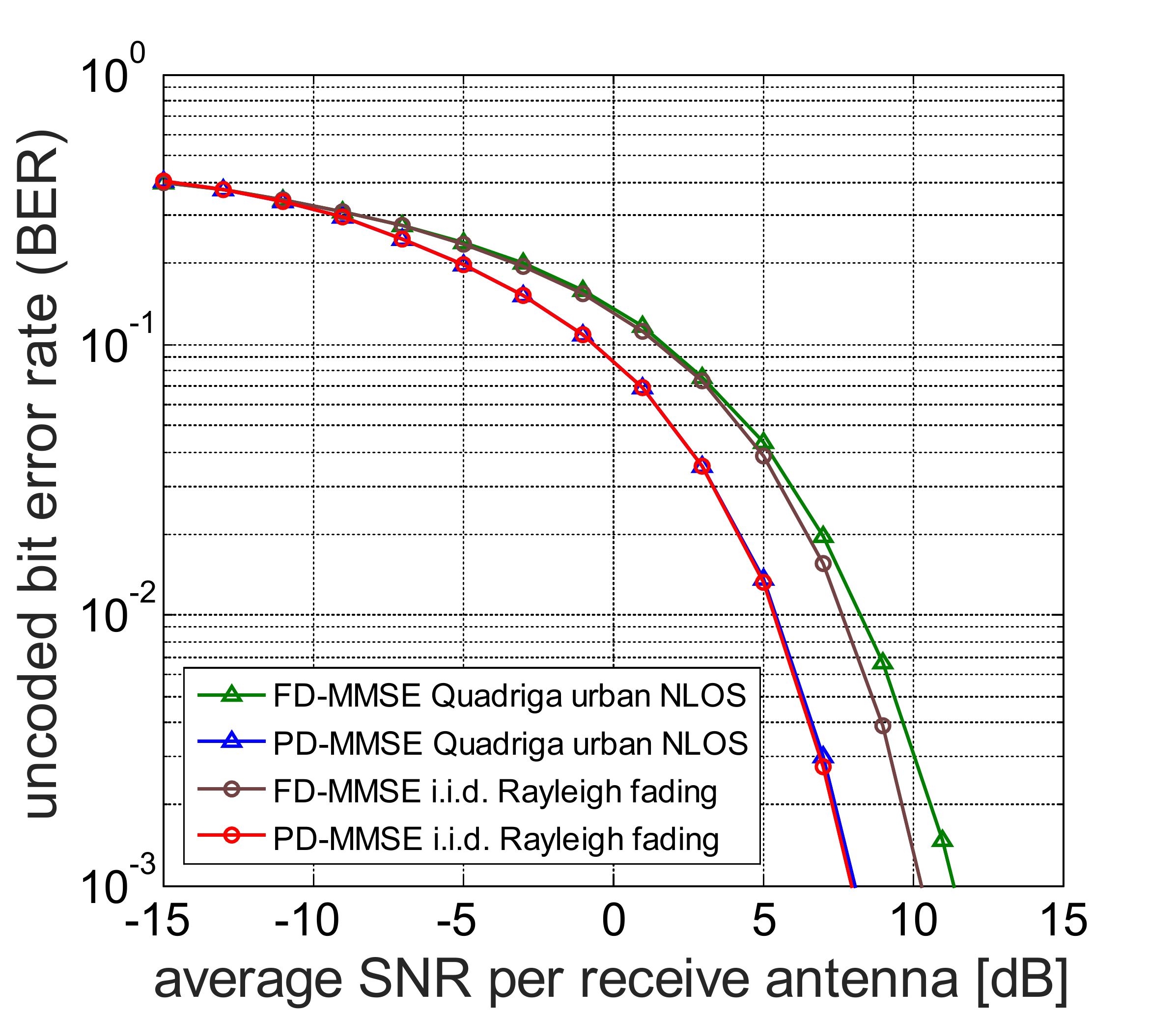}\label{fig:ber}}
\subfigure[Data transfer size vs. $N_\text{coh}$]{\includegraphics[width=0.237\textwidth]{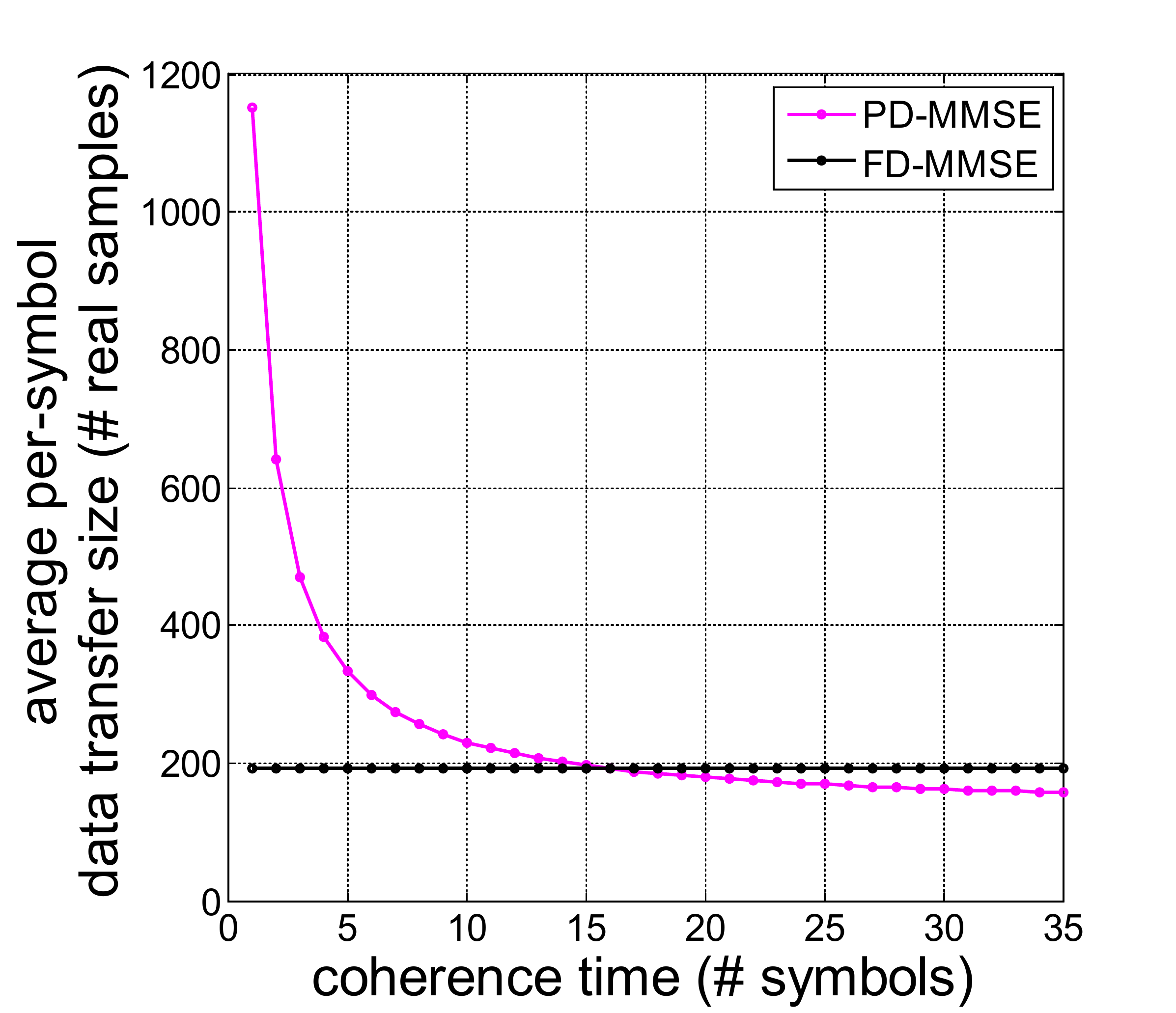}\label{fig:datasize}}
\caption{PD vs.\ FD architecture trade-off on uncoded BER 
and data transfer size.}
\label{fig:pdfdtradeoff}
\end{figure}

\subsection{PD Architecture vs. FD Architecture}
\label{sec:pdvsfd}
We now compare the PD architecture and FD architecture and explore the trade-offs for architecture selection. We first focus on a case study of the linear MMSE data detector in the uplink with the PD and FD architectures, to showcase how architecture selection for a certain algorithm affects the error rate performance, computational complexity, and data transfer size under different antenna configurations and channel conditions, such as channel coherence time. We later extend our analysis to precoding in the downlink and discuss more algorithm variations and their trade-offs in~\fref{sec:tradealg}.   

Linear uplink data detection can be formulated as the following optimization problem (we omit the superscript $^\text{ul}$):
\begin{align} \label{eq:MMSEequalization}
\hat\bmx = \argmin_{\bmx\in\complexset^U} \, \textstyle  \vecnorm{\bmy-\bH\bmx}_2^2 + \rho\vecnorm{\bmx}_2^2.
\end{align}
%
Here, $\rho=\frac{N_0}{E_s}$ is used for linear MMSE equalization where $E_s$ is the per-UE transmit power. The corresponding closed form solution for (centralized) MMSE detection is given by $\hat\bmx=(\bH^H\bH+\rho\bI)^{-1}\bH^H\bmy$. For PD-MMSE detection, we first compute $\bG_c=\bH_c^H\bH_c$ and $\bmy_c^{\text{MRC}}=\bH_c^H\bmy$ at each decentralized cluster. We then fuse the matrices $\bG_c$ and $\bmy_c^{\text{MRC}}$ to form the global Gram matrix  $\bG=\bH^H\bH=\Sigma^C_{c=1}{\bG_c}$ and MRC vector $\bmy^{\text{MRC}}=\bH^H\bmy=\Sigma^C_{c=1}{\bmy_c^{\text{MRC}}}$ at the centralized processor. Finally, we calculate the matrix inversion to perform equalization at the centralized processor, which yields the global equalization output~$\hat\bmx$ and noise variance $\sigma^2$. For FD-MMSE detection, we first compute $\bG_c$ and $\bmy_c^{\text{MRC}}$ locally, the same as for PD-MMSE. We then calculate a matrix inversion and perform equalization locally to obtain the local estimate $\hat{\bmx}_c$ and noise variance $\sigma_c^2$. Finally, we fuse the local estimates $\hat{\bmx}_c$ together with the local noise variances $\sigma_c^2$ at the centralized processor, which yields the global equalization output $\hat\bmx=\Sigma_{c=1}^C{\lambda_c\hat{\bmx}_c}$, where $\lambda_c=\frac{1}{\sigma_c^2}(\Sigma_{c'=1}^C{\frac{1}{\sigma_{c'}^2}})^{-1}$; see~\cite{isit17,tsp19} for the details. 

From the above explanations, we see that the \emph{timing complexity}, which we measure as the number of real-valued multiplications, is approximately the same\footnote{Here, the timing complexity indicates that decentralized computations performed \emph{in parallel} at each cluster should be counted \emph{only once} to reflect the computation latency on real hardware.} for both architectures. The key factor that can lead to different efficiency of PD and FD architectures lies at the data transfer size at the data fusion stage. For the PD architecture, data fusion requires the transfer of both local $\bG_c$, a $U\times U$ complex-value Hermitian matrix consisting of $U^2$ unique real values 
and local $\bmy_c^{\text{MRC}}$, a $U$-dimensional complex-valued vector that contains $2U$ real values, which leads to a total of $C\times (U^2+2U)$ 
real values for each data symbol. If we consider a typical scenario for which the estimated channel in the uplink is static across $N_\text{coh}$ contiguous symbols, then the local Gram matrix $\bG_c$, which depends only on the local channel matrix $\bH_c$, can be transferred only once for every $N_\text{coh}$ symbols, while the transfer of $\bmy_c^{\text{MRC}}$ is required for every symbol. Therefore, the average data transfer size $m_\text{PD}$ for each symbol at the fusion stage of PD architecture is as follows:
\begin{align}
\label{eq:pd}
m_\text{PD}=\frac{C\times (U^2 + 2N_\text{coh}U)}{N_\text{coh}}.
\end{align}
In contrast, for the FD architecture, data fusion requires the transfer of the local $\bmx_c$, a $U$-entry complex-valued vector, and $\sigma_c$, a $U$-entry real-valued vector, which leads to a total of $3U$ real values that must be transferred for every symbol regardless of the channel coherence time. Therefore, the average data transfer size $m_\text{FD}$ for each symbol in FD architecture is
\begin{align}
\label{eq:fd}
m_\text{FD}=\frac{C\times 3N_\text{coh}U}{N_\text{coh}}=3CU.
\end{align}
When extended to a multi-subcarrier transmission, e.g., using orthogonal frequency-division multiplexing (OFDM), the quantities $m_\text{PD}$ and $m_\text{FD}$ represent the average data transfer size per symbol \emph{on each subcarrier}. From~\fref{eq:pd} and \fref{eq:fd}, we see that the channel coherence time determines whether the PD or FD architecture requires more or less data to be transferred. 

\subsection{Trade-off Analysis}
While the data transfer size and efficiency of the PD and FD architectures depend on system parameters and channel conditions, the PD architecture always outperforms the FD architecture in terms of error-rate performance, as it is able to achieve the same performance as centralized data detection~\cite{isit17,tsp19}. 
Figure~\ref{fig:ber} 
compares the uncoded bit error-rate (BER) of PD-MMSE and FD-MMSE data detection for a single-carrier system with a cluster size of $B_c=32$, $C=4$ clusters, and $U=16$ UEs  with 16-QAM. 
We simulate the BER performance with a simple i.i.d.\ Rayleigh fading channel, and
a more realistic urban micro-campus non-line-of-sight (NLOS) Quadriga channel~\cite{quadriga}, where we place the UEs randomly in a 120$^\circ$ sector at a distance of $50$ to $100$  from the BS, which is using a uniform linear array. 

We see that PD-MMSE clearly outperforms FD-MMSE under both channel environments in terms of uncoded BER. 
To showcase the trade-off of PD and FD on BER vs. data transfer size, in~\fref{fig:datasize}, we also compare the average per-symbol data transfer size $m_\text{PD}$ and $m_\text{FD}$ for PD and FD architectures, respectively, under the same antenna configuration but for different channel coherence times characterized by $N_\text{coh}$. We find that at small coherence times $N_\text{coh}$, if $N_\text{coh}<U$, 
then $m_\text{FD}<m_\text{PD}$, which implies that one should select the FD rather than the PD architecture if the data fusion efficiency has higher priority than BER performance.
If $N_\text{coh}>U$, then $m_\text{PD}<m_\text{FD}$, then the PD architecture is always preferred for both better data fusion efficiency and BER performance.

Similarly to the uplink, we can use the PD and FD architectures to realize decentralized precoding in the downlink, e.g., using linear Wiener filter (WF) precoding~\cite{asilomar18}. 
For PD-WF precoding, the total data transfer size, which consists of both fusion of local Gram matrices $\bG_c$ and broadcasting of centralized whitened vector (scaled with $U$), is always larger than that of FD-WF precoding, which only requires the broadcasting of the transmit vector $\bms$ (scaled with $U$) and a scalar power allocation value. 
We can calculate similar average per-symbol data transfer sizes for downlink precoding by $m_\text{PD}=\frac{C(U^2+2N_\text{coh}U)}{N_\text{coh}}$ 
and $m_\text{FD}=\frac{C(1+2N_\text{coh}U)}{N_\text{coh}}$. Clearly, the FD architecture is more efficient on data transfer since $m_\text{FD}<m_\text{PD}$ holds in typical scenarios. 
Only if $N_\text{coh} \gg U^2$, i.e., for nearly static channels, we have $m_\text{FD}\approx m_\text{PD}$. 
However, for the BER performance, PD-WF is never worse than that of FD-WF. 
Therefore, selecting the  PD or FD architecture for the downlink is solely determined by the system designer's preference, i.e., whether BER or data transfer is the critical factor. 

\section{Algorithm Trade-offs}
\label{sec:tradealg}
We now investigate the selection of data detection or precoding algorithms, given that either the PD or FD architecture has been already chosen. We first discuss the trade-offs for various decentralized algorithms, where we focus on \emph{explicit} and \emph{implicit} matrix-inversion-based uplink equalization methods with the PD architecture. A similar  trade-off analysis applies for the FD architecture and downlink systems. 
We also explore situations in which we are able to store and reuse computation results from the uplink in order to facilitate downlink precoding, which enables additional complexity reduction in a time-division duplex~(TDD) system.

\subsection{Explicit vs.\ Implicit Algorithms}
\label{sec:exvsim}
As discussed in~\fref{sec:tradearch}, a closed form solution for (centralized) MMSE data detection is given by $\hat{\bmx}=(\bG+\rho\bI)^{-1}\bmy^{\text{MRC}}$ and requires (i)  Gram matrix computation and (ii) matrix inversion of the regularized Gram matrix, both of which are computationally intensive but only depend on the channel matrix $\bH$. 
For the PD or FD architectures, the timing complexity 
of Gram matrix computation at each cluster is reduced because of smaller matrix size ($B_c\times U$), while the dimension of the complete Gram matrix is still the same (i.e., $U\times U$).
This implies that the matrix inversion timing complexity is the same as that of the centralized MMSE equalizer. 
To reduce complexity, we can (i) take advantage of channel coherence, which allows us to reuse intermediate computation results across $N_\text{coh}$ symbols, and (ii) avoid an explicit computation of the matrix inversion.

Consider PD-MMSE as an example. At each cluster, if we compute $\bG_c$ only once and reuse it across $N_\text{coh}$ receive symbols, then the average timing complexity (number of real-valued multiplications) per receive symbol for computing $\bG_c+\rho\bI$ can be as low as $2B_cU^2/N_\text{coh}$. 
However, the computation of $\bmy_c^{\text{MRC}}=\bH_c^H\bmy_c$ relies not only on channel $\bH_c$ but also on the receive symbol $\bmy_c$, which leads to $4B_cU$ timing complexity per receive symbol. Then for explicit Cholesky-based matrix inversion of a $U\times U$ matrix, 
which only depends on $\bH$, we can still compute the matrix inversion only once and reuse it across $N_\text{coh}$ receive symbols, and therefore the average complexity per receive symbol is $(\frac{10}{3}U^3-\frac{4}{3}U)/N_{coh}$ \cite{Wu_JSPS2017}. 
Finally, to compute the equalization output $\hat\bmx$, the  matrix-vector multiplication requires an additional $4U^2$ operations per receive symbol. Therefore, the timing complexity $n_\text{ex}$ of all above steps to obtain $\hat\bmx$ averaged on each symbol for the \emph{explicit} matrix inversion based PD-MMSE is
\begin{align}
\label{eq:exmmse}
n_\text{ex}= \textstyle \left(2B_cU^2+\frac{10}{3}U^3-\frac{4}{3}U\right)\!/N_\text{coh}+4B_cU+4U^2.
\end{align}
The literature describes a number of implicit methods that can be used for directly computing $\hat\bmx$ while avoiding an explicit matrix inversion, such as the decentralized conjugate gradient method~\cite{Li_Asilomar_16} or decentralized coordinate descent method~\cite{iscas19}. 
Such iterative methods typically obtain an approximate result\footnote{While the conjugate gradient method can obtain the exact MMSE solution after $U$ iterations, we typically perform a smaller number of iterations (less than $U$) to reduce complexity.}, which entails a small BER loss. Furthermore, such implicit methods are unable to exploit the benefits of channel coherence since all iterative updates need to be computed for every symbol. 
We therefore propose to integrate \emph{implicit} Cholesky-based MMSE detection~\cite{Wu_JSPS2017} with the PD architecture, which is not only able to compute the exact linear MMSE equalizer but also realize low complexity compared to other iterative methods, especially for large $N_{coh}$ when the intermediate Cholesky decomposition results can be reused. 
The \emph{implicit} PD-MMSE algorithm computes the local Gram matrix $\bG_c$ and MRC vector $\bmy_c^{\text{MRC}}$ and fuses them to obtain the global Gram matrix $\bG$ and MRC vector $\bmy^{\text{MRC}}$ similarly to explicit PD-MMSE. After fusion, the implicit PD-MMSE method factorizes the regularized Gram matrix $\bA=\bG+\rho\bI$ by Cholesky decomposition $\bA=\bL\bL^H$, where $\bL$ represents a lower triangular matrix. One can then solve $\bL\bmz=\bmy^{\text{MRC}}$ and finally $\bL^H\hat{\bmx}=\bmz$, by forward and backward substitution, respectively, in order to obtain $\hat\bmx$. While the forward and backward substitutions have to be carried out for every symbol with a total of $4U^2$ operations (including real-valued multiplications and divisions), the Cholesky decomposition, which dominates the complexity at $\frac{2}{3}U^3-\frac{2}{3}U$, can be computed only once every $N_{coh}$ symbols. Therefore, the resulting timing complexity $n_{im}$ per symbol for the implicit Cholesky based PD-MMSE approach is
\begin{align}
\label{eq:immmse}
n_\text{im} = \textstyle \left(2B_cU^2+\frac{2}{3}U^3-\frac{2}{3}U\right)\!/N_\text{coh}+4B_cU+4U^2.
\end{align}
In OFDM systems, the quantities $n_\text{ex}$ and $n_\text{im}$ represent the detection complexity on each subcarrier. 

\begin{figure}[tp]
\centering
\includegraphics[width=0.6\columnwidth]{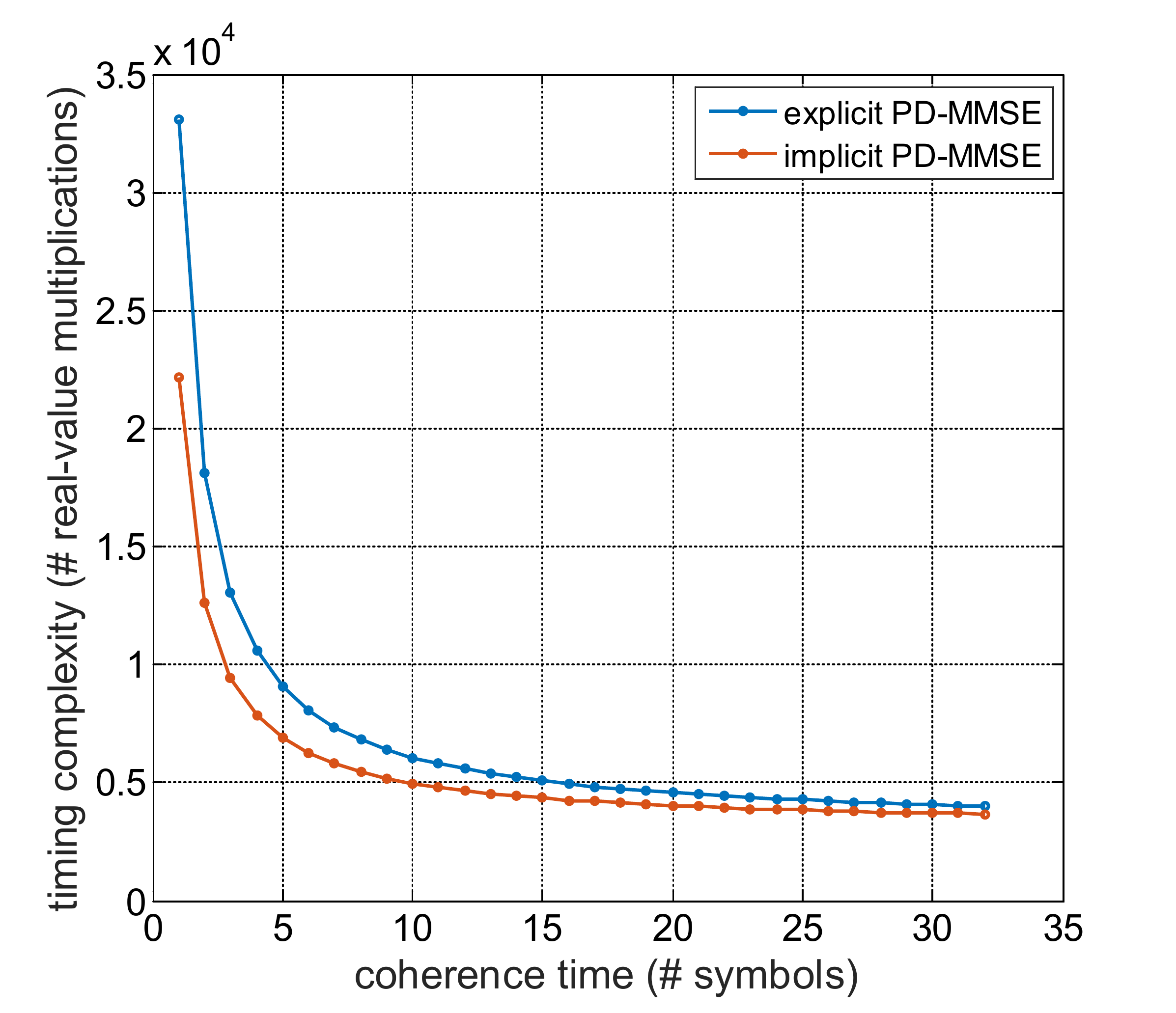}
\caption{Complexity of explicit vs. implicit PD-MMSE at different $N_\text{coh}$.} 
\label{fig:exvsim}
\end{figure}

By comparing \fref{eq:exmmse} and \fref{eq:immmse}, we see that the $U^3$ term in $n_\text{ex}$ and $n_\text{im}$ dominates the complexity and $n_\text{im} < n_\text{ex}$ since $n_\text{im}$ has smaller constant associated with this term; this indicates that the implicit method can reduce complexity at \emph{no loss} in terms of  BER compared to the explicit method. 
Figure~\ref{fig:exvsim} shows an example for  a system with $B_c=32$, $U=16$, and $C=4$ where we compare the timing complexity of explicit PD-MMSE and implicit PD-MMSE depending on the coherence time. We observe that the implicit method always achieves lower complexity, whereas the complexity is similar to that of the explicit method for large values of $N_\text{coh}$.

\begin{figure*}[t]
\centering
\subfigure[BER: PD-MMSE detection vs. PD-ZF detection]{\includegraphics[width=0.32\textwidth]{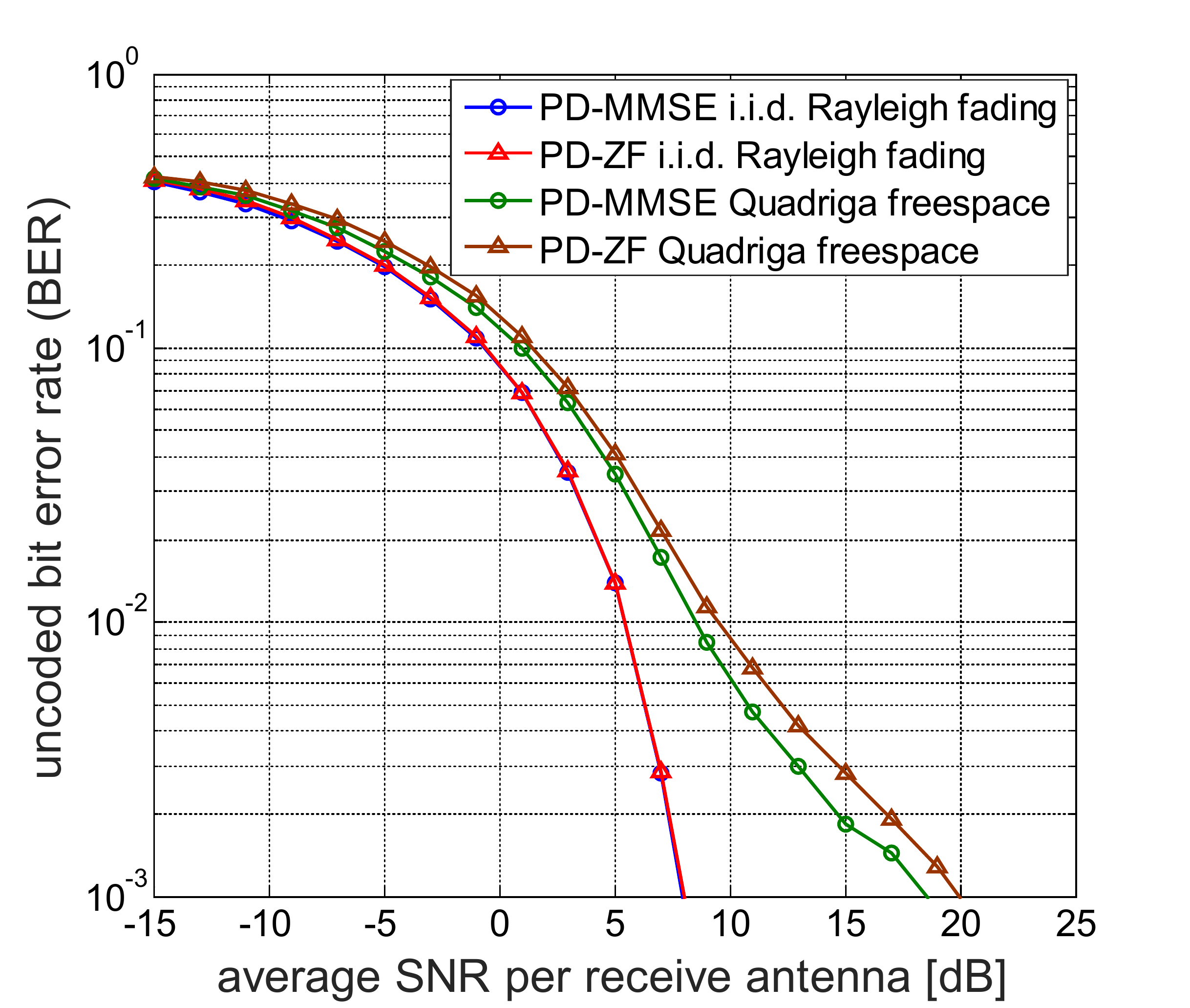}\label{fig:ulber}}
\subfigure[BER: PD-WF precoding vs. PD-ZF precoding]{\includegraphics[width=0.32\textwidth]{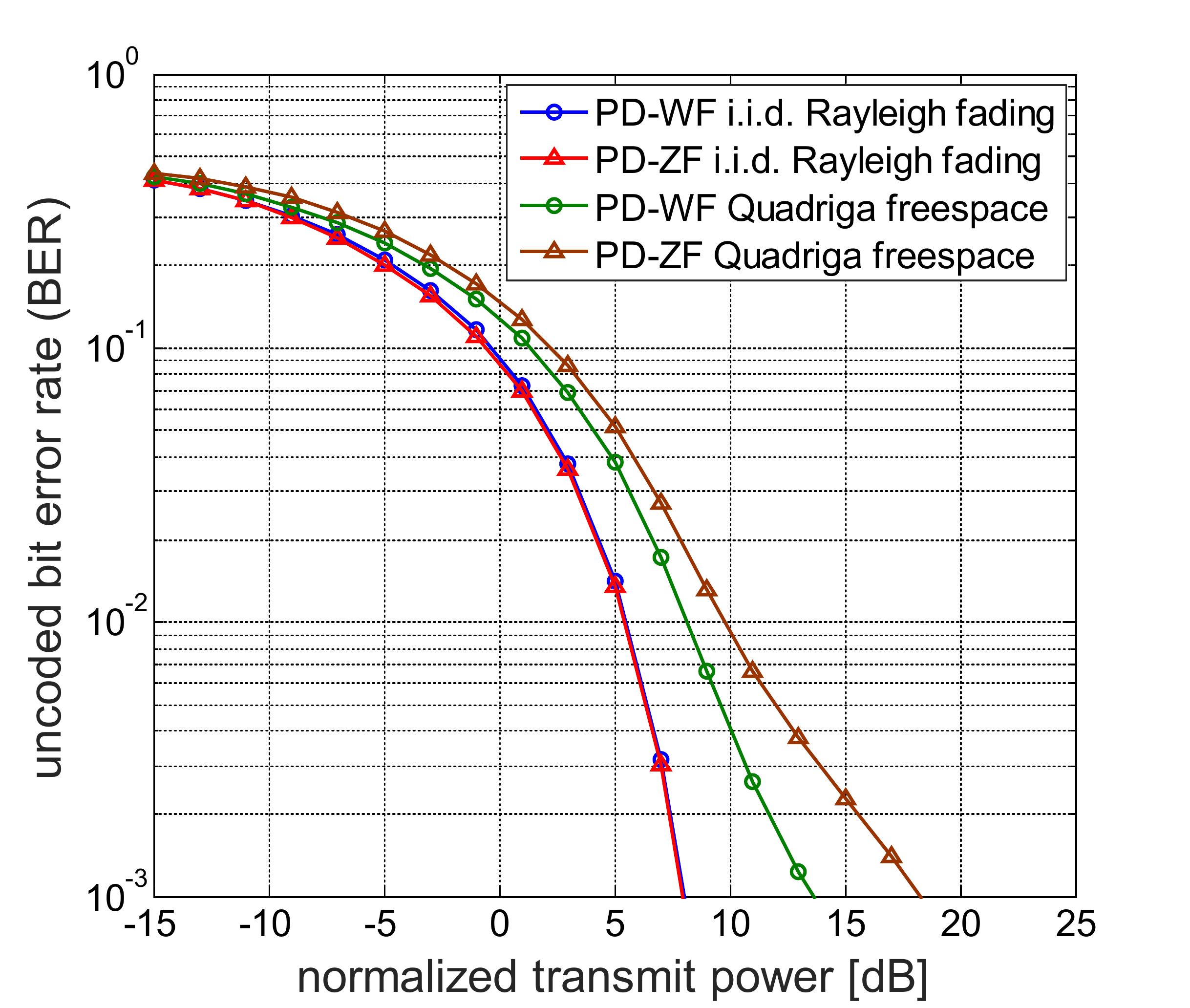}\label{fig:dlber}}
\subfigure[Complexity of integrated UL and DL]{\includegraphics[width=0.32\textwidth]{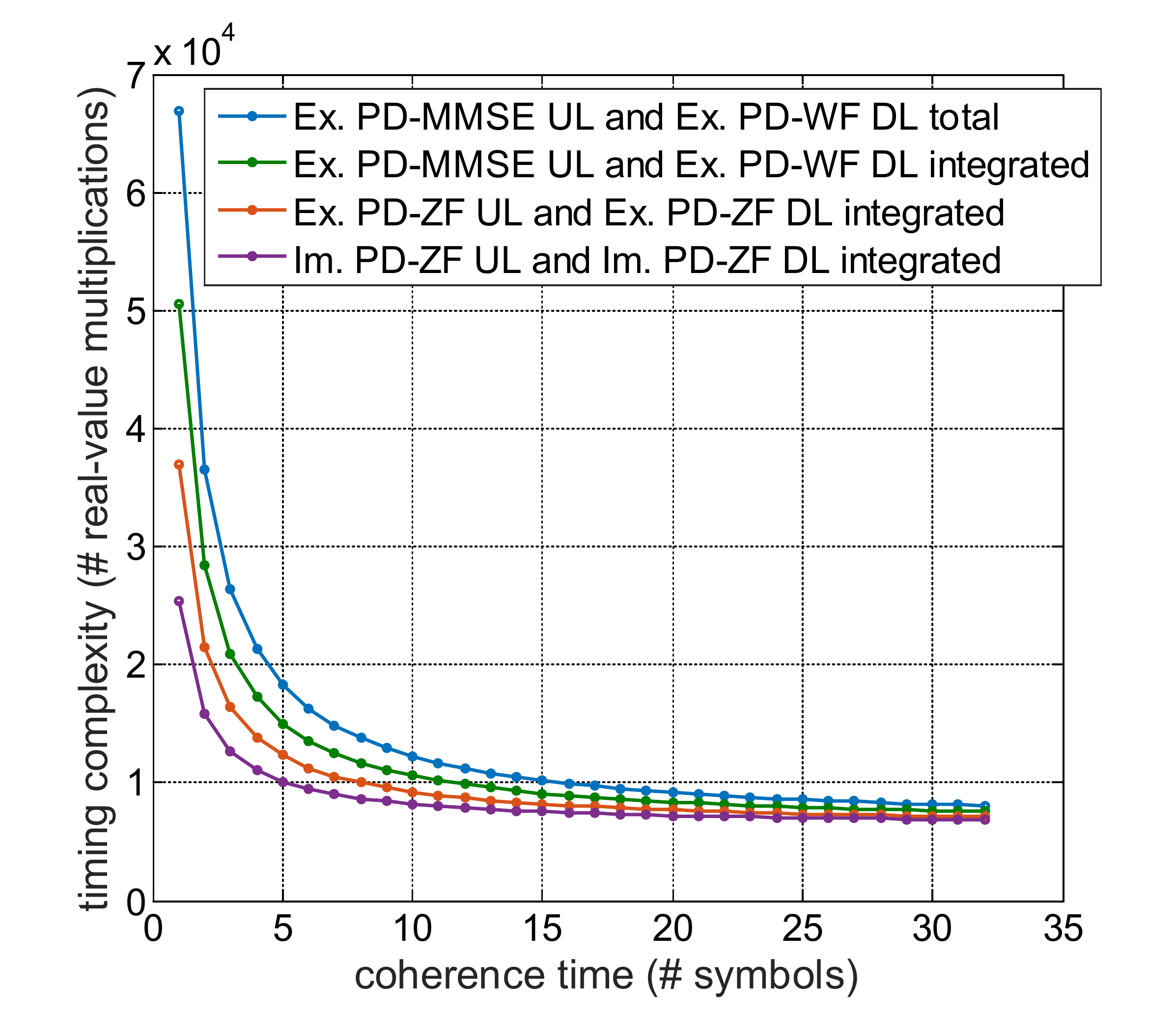}\label{fig:compreduce}}
\caption{UL and DL integration trade-offs on BER and complexity for a system with $C=4, U=16, B_c=32$. (a) and (b) show the BER performance comparison of PD-MMSE vs. PD-ZF in uplink, and PD-WF vs. PF-ZF in downlink, respectively, at 16QAM modulation. PD-ZF sacrifices the BER performance, especially under a more realistic Quadriga channel, while achieving more significant complexity reduction with UL and DL integration, as shown in (c). We use the \emph{total} complexity of individually computed (without UL and DL integration) explicit PD-MMSE and PD-WF as the baseline (the blue curve) for comparison.}
\label{fig:integratetradeoff}
\end{figure*}


We conclude by noting that we can similarly leverage both channel coherence and implicit inversion to reduce the complexity for FD-based uplink data detection. 
However, for the downlink, implicit inversion is not particularly helpful for the PD-WF or FD-WF precoding algorithms, which require the computation of an optimal scaling factor that depends on the explicit matrix inversion result of a regularized Gram matrix~\cite{asilomar18}. 

As an alternative, one can resort to ZF-based algorithms (e.g., PD-ZF and FD-ZF), which enable implicit methods for both uplink detection and downlink precoding. Here, for ZF precoding, we can simply scale the normalized ZF precoding vector $\frac{\hat{\bmx}}{{\lVert\hat{\bmx}\rVert}_2}$ 
without the need of explicit matrix inversion to satisfy the same transmit power constraint as WF precoding. 

\begin{figure*}[tp]
\centering
\includegraphics[width=2.0\columnwidth]{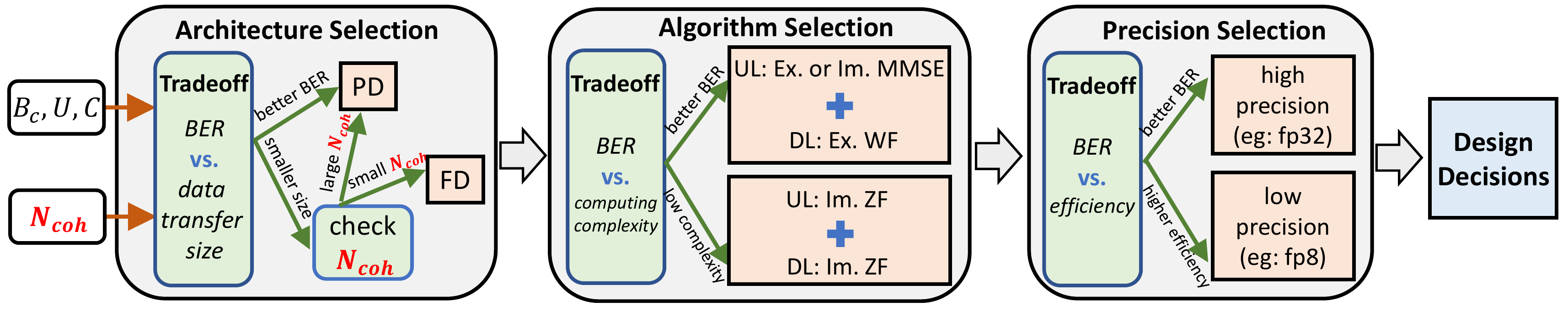}
\caption{Practical design decision flow: DBP architecture, algorithm and precision trade-off and selection.
} 
\label{fig:pipeline}
\end{figure*}

\begin{figure}[tp]
\centering
\includegraphics[width=0.8\columnwidth]{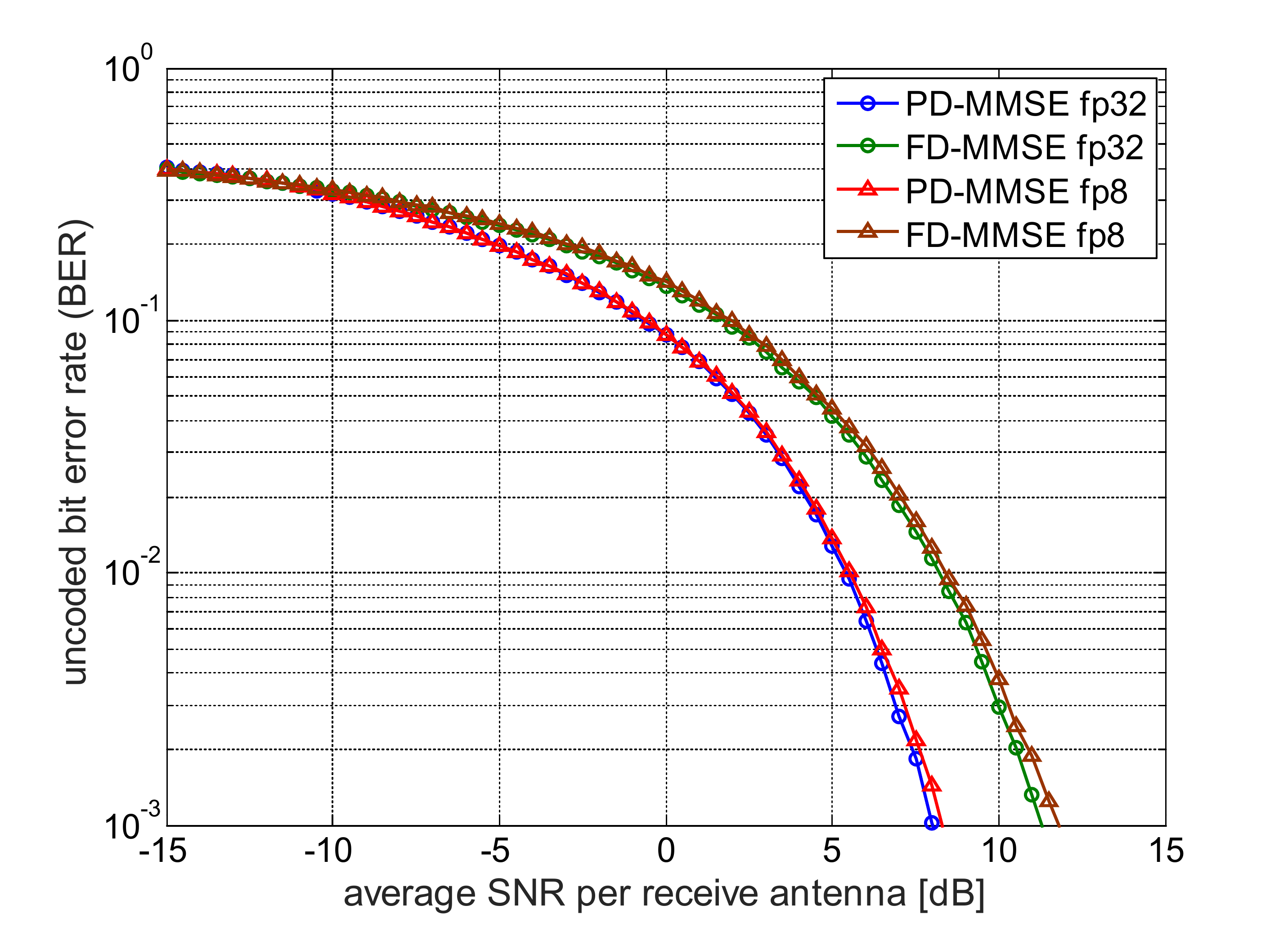}

\caption{BER comparison: fp32 vs. fp8 for PD-MMSE and FD-MMSE 
} 
\label{fig:quant}

\end{figure}

\subsection{Reusing Uplink Results for Downlink}
\label{sec:reuseul}
Due to channel reciprocity in TDD systems, we have $\bH^\text{dl}=(\bH^\text{ul})^T$, and therefore $\bG^\text{dl}={\bH^\text{dl}}(\bH^\text{dl})^H=(\bH^\text{ul})^T(\bH^\text{ul})^*=((\bH^\text{ul})^H\bH^\text{ul})^*=(\bG^\text{ul})^*=(\bG^\text{ul})^T$, where $(\cdot)^*$ indicates entry-wise complex conjugate operation and $\bG^\text{ul}$ is an Hermitian matrix. 
This indicates that the Gram matrix computed in the uplink, can also be reused in the downlink by a simple transpose. We can also take advantage of channel reciprocity under decentralized architectures. For example, in FD-MMSE, the local Gram matrix $\bG_c^\text{ul}$ is computed. If we perform FD-WF precoding in the downlink, and each cluster estimates the local precoding vector $\hat{\bmx}_c=\frac{1}{\beta_c}(\bH_c^\text{dl})^H(\bG_c^\text{dl}+\kappa_c\bI)^{-1}\bms$, where $\frac{1}{\beta_c}$ is the scaling factor to satisfy the transmit power constraint and $\kappa_c$ regularizes $\bG_c^\text{dl}$ as detailed in~\cite{asilomar18}, then we can store the uplink matrix $\bG_c^\text{ul}$ and reuse it as $\bG_c^\text{dl}=(\bG_c^\text{ul})^T$ for FD-WF precoding, even across $N_\text{coh}$ symbols. This approach yields an additional  $2B_cU^2/N_\text{coh}$ complexity reduction per symbol for a system that integrates FD-MMSE detection and FD-WF precoding by re-using intermediate results compared to the total complexity of computing them individually. A similar approach can be used for PD-MMSE detection and PD-WF precoding: in PD-MMSE detection, we aggregate local $\bG_c^{\text{ul}}$ and compute the global $\bG^{\text{ul}}$ at the centralized node, and transpose it for PD-WF precoding.

If we consider ZF detection and precoding, where the Gram regularization coefficients $\rho=0$ and $\kappa=0$, respectively, we can even store the \emph{matrix inversion} result $(\bG^{\text{ul}})^{-1}$ computed in ZF detection and reuse it for ZF precoding which requires $(\bG^{\text{dl}})^{-1}=((\bG^{\text{ul}})^T)^{-1}=((\bG^{\text{ul}})^{-1})^T$. When $(\bG^{\text{ul}})^{-1}$ is computed explicitly, we can also use it to compute the scaling factor $\beta$ for ZF precoding in a similar way like WF precoding~\cite{asilomar18}. When it is computed implicitly, we should store and reuse the Cholesky decomposition result $\bG^{\text{ul}}=\bL\bL^H$ rather than the matrix inversion result for implicit ZF precoding which relies on $\bG^{\text{dl}}=(\bL\bL^H)^T=\bL^*(\bL^*)^H$, and finally scale the normalized ZF precoding vector $\frac{\hat{\bmx}}{{\lVert\hat{\bmx}\rVert}_2}$ 
to reach the power constraint. Similarly, under decentralized scenarios (PD or FD architecture), for example, with an integrated pipeline of PD-ZF detection and PD-ZF precoding, we can reuse such explicit or implicit inversion results to realize further complexity reduction compared to the integrated pipeline of PD-MMSE detection and PD-WF precoding, at the cost of BER performance degradation. 

As an example, Fig.~\ref{fig:ulber} and~\ref{fig:dlber} compare the uncoded BER performance of PD-MMSE vs. PD-ZF for uplink detection, and PD-WF vs. PD-ZF for downlink precoding, respectively, in a single-carrier system. We see that PD-ZF methods entail BER performance loss compared to PD-MMSE and PD-WF, expecially under realistic Quadriga freespace channel~\cite{quadriga}. Fig.~\ref{fig:compreduce} compares the timing complexity of different pipelines of uplink detection and downlink precoding at different $N_{coh}$. Using the total complexity of individually computed explicit PD-MMSE detection and PD-WF precoding as the baseline, we show that the integration of PD-MMSE and PD-WF by reusing $\bG^{\text{ul}}$ effectively reduces complexity, and the integration of PD-ZF detection and PD-ZF precoding achieves further complexity reduction as expected, especially when incorporated with implicit methods. When $N_\text{coh}$ increases, the difference among those complexity curves decreases, indicating that the channel coherence plays a more important role on complexity reduction than channel reciprocity at a large $N_\text{coh}$, while at a small $N_\text{coh}$, exploiting channel reciprocity is more critical.

\section{Data Precision Trade-offs}
\label{sec:tradeprec}

%
Data precision is another factor in the design space. Given a decentralized architecture and algorithm, reducing the data precision can improve efficiency on modern computing fabrics 
due to fewer compiled machine instructions and memory transactions, and smaller inter-cluster data transfer sizes. 
For example, when using 8-bit floating point (fp8), we can pack four fp8 values into a fp32 value and execute a 
\emph{vectorized} computation instruction (such as vectorized addition, multiplication, etc) in a single-instruction-multiple-data (SIMD) manner to process four fp8 values in parallel within the single instruction, contributing to $4\times$ smaller number of  instructions and memory transactions on the processor, and also reducing the inter-cluster bandwidth requirement by $4\times$ compared to corresponding fp32 design. However, low precision sacrifices numerical accuracy and thus reduces the BER performance of corresponding detection or precoding algorithms. Fig.~\ref{fig:quant} compares the uncoded BER performance at fp32 vs. fp8 precision for PD-MMSE and FD-MMSE detectors under a realistic NLOS Quadriga channel~\cite{quadriga} with system configurations of $C=4$, $U=16$, $B_c=32$ and 16-QAM modulation. Here, an fp8 value contains 1 sign bit, 2 mantissa bits and 5 exponent bits while an fp32 value contains 1 sign bit, 23 mantissa bits and 8 exponent bits. We see in Fig.~\ref{fig:quant} that fp8 precision only entails a small BER performance loss compared to fp32 precision. In practice, the selection of data precision depends on the trade-off of BER performance vs. efficiency given certain system and environment configurations.



\section{Conclusions}
\label{sec:conclusions}
%
We have discussed the design trade-offs across architecture, algorithm, and data precision levels for decentralized baseband processing (DBP) in massive MU-MIMO systems, and proposed a practical design flow that jointly considers critical metrics for DBP including computational complexity, data transfer sizes, and error-rate performance. As summarized in Fig.\ref{fig:pipeline}, given certain system configurations and channel conditions, one should first select the PD or FD architecture by trading off BER vs. data transfer size, and then decide on the detection and precoding algorithms with the selected architecture according to the  BER vs.\ complexity trade-off. Finally, one can the lower numerical precision if higher efficiency is more important than BER performance. To realize minimal computational complexity and data transfer size at no or little loss of BER, we have provided insights on taking advantage of both channel reciprocity and channel coherence properties by reusing intermediate results. In the future, we expect to build reconfigurable massive MU-MIMO software-defined radios based on programmable computing fabrics, such as GPUs or FPGAs, in order to dynamically adapt to time-varying system parameters and realize effective trade-offs.

%
\bibliographystyle{IEEEbib}
\bibliography{MIMO}
\balance


\end{document}

%% file: vmr-symbols-vecbold.tex
\usepackage{amssymb}
\usepackage{amsfonts}
\usepackage{mathrsfs}
\usepackage{xspace}
\usepackage{bm}
\usepackage{upgreek}

\newcommand{\safemath}[2]{\newcommand{#1}{\ensuremath{#2}\xspace}}

\safemath{\bma}{\mathbf{a}}
\safemath{\bmb}{\mathbf{b}}
\safemath{\bmc}{\mathbf{c}}
\safemath{\bmd}{\mathbf{d}}
\safemath{\bme}{\mathbf{e}}
\safemath{\bmf}{\mathbf{f}}
\safemath{\bmg}{\mathbf{g}}
\safemath{\bmh}{\mathbf{h}}
\safemath{\bmi}{\mathbf{i}}
\safemath{\bmj}{\mathbf{j}}
\safemath{\bmk}{\mathbf{k}}
\safemath{\bml}{\mathbf{l}}
\safemath{\bmm}{\mathbf{m}}
\safemath{\bmn}{\mathbf{n}}
\safemath{\bmo}{\mathbf{o}}
\safemath{\bmp}{\mathbf{p}}
\safemath{\bmq}{\mathbf{q}}
\safemath{\bmr}{\mathbf{r}}
\safemath{\bms}{\mathbf{s}}
\safemath{\bmt}{\mathbf{t}}
\safemath{\bmu}{\mathbf{u}}
\safemath{\bmv}{\mathbf{v}}
\safemath{\bmw}{\mathbf{w}}
\safemath{\bmx}{\mathbf{x}}
\safemath{\bmy}{\mathbf{y}}
\safemath{\bmz}{\mathbf{z}}
\safemath{\bmzero}{\mathbf{0}}
\safemath{\bmone}{\mathbf{1}}

\bmdefine{\biad}{a}
\bmdefine{\bibd}{b}
\bmdefine{\bicd}{c}
\bmdefine{\bidd}{d}
\bmdefine{\bied}{e}
\bmdefine{\bifd}{f}
\bmdefine{\bigd}{g}
\bmdefine{\bihd}{h}
\bmdefine{\biid}{i}
\bmdefine{\bijd}{j}
\bmdefine{\bikd}{k}
\bmdefine{\bild}{l}
\bmdefine{\bimd}{m}
\bmdefine{\bind}{n}
\bmdefine{\biod}{o}
\bmdefine{\bipd}{p}
\bmdefine{\biqd}{q}
\bmdefine{\bird}{r}
\bmdefine{\bisd}{s}
\bmdefine{\bitd}{t}
\bmdefine{\biud}{u}
\bmdefine{\bivd}{v}
\bmdefine{\biwd}{w}
\bmdefine{\bixd}{x}
\bmdefine{\biyd}{y}
\bmdefine{\bizd}{z}

\bmdefine{\bixid}{\xi}
\bmdefine{\bilambdad}{\lambda}
\bmdefine{\bimud}{\mu}
\bmdefine{\bithetad}{\theta}
\bmdefine{\biphid}{\phi}
\bmdefine{\bideltad}{\delta}

\safemath{\bmia}{\biad}
\safemath{\bmib}{\bibd}
\safemath{\bmic}{\bicd}
\safemath{\bmid}{\bidd}
\safemath{\bmie}{\bied}
\safemath{\bmif}{\bifd}
\safemath{\bmig}{\bigd}
\safemath{\bmih}{\bihd}
\safemath{\bmii}{\biid}
\safemath{\bmij}{\bijd}
\safemath{\bmik}{\bikd}
\safemath{\bmil}{\bild}
\safemath{\bmim}{\bimd}
\safemath{\bmin}{\bind}
\safemath{\bmio}{\biod}
\safemath{\bmip}{\bipd}
\safemath{\bmiq}{\biqd}
\safemath{\bmir}{\bird}
\safemath{\bmis}{\bisd}
\safemath{\bmit}{\bitd}
\safemath{\bmiu}{\biud}
\safemath{\bmiv}{\bivd}
\safemath{\bmiw}{\biwd}
\safemath{\bmix}{\bixd}
\safemath{\bmiy}{\biyd}
\safemath{\bmiz}{\bizd}

\safemath{\bmxi}{\bixid}
\safemath{\bmlambda}{\bilambdad}
\safemath{\bmmu}{\bimud}
\safemath{\bmtheta}{\bithetad}
\safemath{\bmphi}{\biphid}
\safemath{\bmdelta}{\bideltad}

\safemath{\bA}{\mathbf{A}}
\safemath{\bB}{\mathbf{B}}
\safemath{\bC}{\mathbf{C}}
\safemath{\bD}{\mathbf{D}}
\safemath{\bE}{\mathbf{E}}
\safemath{\bF}{\mathbf{F}}
\safemath{\bG}{\mathbf{G}}
\safemath{\bH}{\mathbf{H}}
\safemath{\bI}{\mathbf{I}}
\safemath{\bJ}{\mathbf{J}}
\safemath{\bK}{\mathbf{K}}
\safemath{\bL}{\mathbf{L}}
\safemath{\bM}{\mathbf{M}}
\safemath{\bN}{\mathbf{N}}
\safemath{\bO}{\mathbf{O}}
\safemath{\bP}{\mathbf{P}}
\safemath{\bQ}{\mathbf{Q}}
\safemath{\bR}{\mathbf{R}}
\safemath{\bS}{\mathbf{S}}
\safemath{\bT}{\mathbf{T}}
\safemath{\bU}{\mathbf{U}}
\safemath{\bV}{\mathbf{V}}
\safemath{\bW}{\mathbf{W}}
\safemath{\bX}{\mathbf{X}}
\safemath{\bY}{\mathbf{Y}}
\safemath{\bZ}{\mathbf{Z}}

\safemath{\bZero}{\mathbf{0}}
\safemath{\bOne}{\mathbf{1}}
\safemath{\bDelta}{\mathbf{\Delta}}
\safemath{\bLambda}{\mathbf{\UpLambda}}
\safemath{\bPhi}{\mathbf{\Upphi}}
\safemath{\bSigma}{\mathbf{\Upsigma}}
\safemath{\bOmega}{\mathbf{\Upomega}}
\safemath{\bTheta}{\mathbf{\Uptheta}}

\bmdefine{\biAd}{A}
\bmdefine{\biBd}{B}
\bmdefine{\biCd}{C}
\bmdefine{\biDd}{D}
\bmdefine{\biEd}{E}
\bmdefine{\biFd}{F}
\bmdefine{\biGd}{G}
\bmdefine{\biHd}{H}
\bmdefine{\biId}{I}
\bmdefine{\biJd}{J}
\bmdefine{\biKd}{K}
\bmdefine{\biLd}{L}
\bmdefine{\biMd}{M}
\bmdefine{\biOd}{N}
\bmdefine{\biPd}{O}
\bmdefine{\biQd}{P}
\bmdefine{\biRd}{R}
\bmdefine{\biSd}{S}
\bmdefine{\biTd}{T}
\bmdefine{\biUd}{U}
\bmdefine{\biVd}{V}
\bmdefine{\biWd}{W}
\bmdefine{\biXd}{X}
\bmdefine{\biYd}{Y}
\bmdefine{\biZd}{Z}

\bmdefine{\biDelta}{\Delta}
\bmdefine{\biLambda}{\Lambda}
\bmdefine{\biPhi}{\Phi}
\bmdefine{\biSigma}{\Sigma}
\bmdefine{\biOmega}{\Omega}
\bmdefine{\biTheta}{\Theta}

\safemath{\bimA}{\biAd}
\safemath{\bimB}{\biBd}
\safemath{\bimC}{\biCd}
\safemath{\bimD}{\biDd}
\safemath{\bimE}{\biEd}
\safemath{\bimF}{\biFd}
\safemath{\bimG}{\biGd}
\safemath{\bimH}{\biHd}
\safemath{\bimI}{\biId}
\safemath{\bimJ}{\biJd}
\safemath{\bimK}{\biKd}
\safemath{\bimL}{\biLd}
\safemath{\bimM}{\biMd}
\safemath{\bimN}{\biNd}
\safemath{\bimO}{\biOd}
\safemath{\bimP}{\biPd}
\safemath{\bimQ}{\biQd}
\safemath{\bimR}{\biRd}
\safemath{\bimS}{\biSd}
\safemath{\bimT}{\biTd}
\safemath{\bimU}{\biUd}
\safemath{\bimV}{\biVd}
\safemath{\bimW}{\biWd}
\safemath{\bimX}{\biXd}
\safemath{\bimY}{\biYd}
\safemath{\bimZ}{\biZd}

\safemath{\bimDelta}{\biDelta}
\safemath{\bimLambda}{\biLambda}
\safemath{\bimPhi}{\biPhi}
\safemath{\bimSigma}{\biSigma}
\safemath{\bimOmega}{\biOmega}
\safemath{\bimTheta}{\biTheta}

\safemath{\setA}{\mathcal{A}}
\safemath{\setB}{\mathcal{B}}
\safemath{\setC}{\mathcal{C}}
\safemath{\setD}{\mathcal{D}}
\safemath{\setE}{\mathcal{E}}
\safemath{\setF}{\mathcal{F}}
\safemath{\setG}{\mathcal{G}}
\safemath{\setH}{\mathcal{H}}
\safemath{\setI}{\mathcal{I}}
\safemath{\setJ}{\mathcal{J}}
\safemath{\setK}{\mathcal{K}}
\safemath{\setL}{\mathcal{L}}
\safemath{\setM}{\mathcal{M}}
\safemath{\setN}{\mathcal{N}}
\safemath{\setO}{\mathcal{O}}
\safemath{\setP}{\mathcal{P}}
\safemath{\setQ}{\mathcal{Q}}
\safemath{\setR}{\mathcal{R}}
\safemath{\setS}{\mathcal{S}}
\safemath{\setT}{\mathcal{T}}
\safemath{\setU}{\mathcal{U}}
\safemath{\setV}{\mathcal{V}}
\safemath{\setW}{\mathcal{W}}
\safemath{\setX}{\mathcal{X}}
\safemath{\setY}{\mathcal{Y}}
\safemath{\setZ}{\mathcal{Z}}
\safemath{\emptySet}{\varnothing}

\safemath{\colA}{\mathscr{A}}
\safemath{\colB}{\mathscr{B}}
\safemath{\colC}{\mathscr{C}}
\safemath{\colD}{\mathscr{D}}
\safemath{\colE}{\mathscr{E}}
\safemath{\colF}{\mathscr{F}}
\safemath{\colG}{\mathscr{G}}
\safemath{\colH}{\mathscr{H}}
\safemath{\colI}{\mathscr{I}}
\safemath{\colJ}{\mathscr{J}}
\safemath{\colK}{\mathscr{K}}
\safemath{\colL}{\mathscr{L}}
\safemath{\colM}{\mathscr{M}}
\safemath{\colN}{\mathscr{N}}
\safemath{\colO}{\mathscr{O}}
\safemath{\colP}{\mathscr{P}}
\safemath{\colQ}{\mathscr{Q}}
\safemath{\colR}{\mathscr{R}}
\safemath{\colS}{\mathscr{S}}
\safemath{\colT}{\mathscr{T}}
\safemath{\colU}{\mathscr{U}}
\safemath{\colV}{\mathscr{V}}
\safemath{\colW}{\mathscr{W}}
\safemath{\colX}{\mathscr{X}}
\safemath{\colY}{\mathscr{Y}}
\safemath{\colZ}{\mathscr{Z}}

\safemath{\opA}{\mathbb{A}}
\safemath{\opB}{\mathbb{B}}
\safemath{\opC}{\mathbb{C}}
\safemath{\opD}{\mathbb{D}}
\safemath{\opE}{\mathbb{E}}
\safemath{\opF}{\mathbb{F}}
\safemath{\opG}{\mathbb{G}}
\safemath{\opH}{\mathbb{H}}
\safemath{\opI}{\mathbb{I}}
\safemath{\opJ}{\mathbb{J}}
\safemath{\opK}{\mathbb{K}}
\safemath{\opL}{\mathbb{L}}
\safemath{\opM}{\mathbb{M}}
\safemath{\opN}{\mathbb{N}}
\safemath{\opO}{\mathbb{O}}
\safemath{\opP}{\mathbb{P}}
\safemath{\opQ}{\mathbb{Q}}
\safemath{\opR}{\mathbb{R}}
\safemath{\opS}{\mathbb{S}}
\safemath{\opT}{\mathbb{T}}
\safemath{\opU}{\mathbb{U}}
\safemath{\opV}{\mathbb{V}}
\safemath{\opW}{\mathbb{W}}
\safemath{\opX}{\mathbb{X}}
\safemath{\opY}{\mathbb{Y}}
\safemath{\opZ}{\mathbb{Z}}
\safemath{\opZero}{\mathbb{O}}
\safemath{\identityop}{\opI}

\safemath{\veca}{\bma}
\safemath{\vecb}{\bmb}
\safemath{\vecc}{\bmc}
\safemath{\vecd}{\bmd}
\safemath{\vece}{\bme}
\safemath{\vecf}{\bmf}
\safemath{\vecg}{\bmg}
\safemath{\vech}{\bmh}
\safemath{\veci}{\bmi}
\safemath{\vecj}{\bmj}
\safemath{\veck}{\bmk}
\safemath{\vecl}{\bml}
\safemath{\vecm}{\bmm}
\safemath{\vecn}{\bmn}
\safemath{\veco}{\bmo}
\safemath{\vecp}{\bmp}
\safemath{\vecq}{\bmq}
\safemath{\vecr}{\bmr}
\safemath{\vecs}{\bms}
\safemath{\vect}{\bmt}
\safemath{\vecu}{\bmu}
\safemath{\vecv}{\bmv}
\safemath{\vecw}{\bmw}
\safemath{\vecx}{\bmx}
\safemath{\vecy}{\bmy}
\safemath{\vecz}{\bmz}

\safemath{\veczero}{\bmzero}
\safemath{\vecone}{\bmone}
\safemath{\vecxi}{\bmxi}
\safemath{\veclambda}{\bmlambda}
\safemath{\vecmu}{\bmmu}
\safemath{\vectheta}{\bmtheta}
\safemath{\vecphi}{\bmphi}
\safemath{\vecdelta}{\bmdelta}

\safemath{\matA}{\bA}
\safemath{\matB}{\bB}
\safemath{\matC}{\bC}
\safemath{\matD}{\bD}
\safemath{\matE}{\bE}
\safemath{\matF}{\bF}
\safemath{\matG}{\bG}
\safemath{\matH}{\bH}
\safemath{\matI}{\bI}
\safemath{\matJ}{\bJ}
\safemath{\matK}{\bK}
\safemath{\matL}{\bL}
\safemath{\matM}{\bM}
\safemath{\matN}{\bN}
\safemath{\matO}{\bO}
\safemath{\matP}{\bP}
\safemath{\matQ}{\bQ}
\safemath{\matR}{\bR}
\safemath{\matS}{\bS}
\safemath{\matT}{\bT}
\safemath{\matU}{\bU}
\safemath{\matV}{\bV}
\safemath{\matW}{\bW}
\safemath{\matX}{\bX}
\safemath{\matY}{\bY}
\safemath{\matZ}{\bZ}
\safemath{\matzero}{\bmzero}

\safemath{\matDelta}{\bDelta}
\safemath{\matLambda}{\bLambda}
\safemath{\matPhi}{\bPhi}
\safemath{\matSigma}{\bSigma}
\safemath{\matOmega}{\bOmega}
\safemath{\matTheta}{\bTheta}

\safemath{\matidentity}{\matI}
\safemath{\matone}{\matO}

\safemath{\rnda}{A}
\safemath{\rndb}{B}
\safemath{\rndc}{C}
\safemath{\rndd}{D}
\safemath{\rnde}{E}
\safemath{\rndf}{F}
\safemath{\rndg}{G}
\safemath{\rndh}{H}
\safemath{\rndi}{I}
\safemath{\rndj}{J}
\safemath{\rndk}{K}
\safemath{\rndl}{L}
\safemath{\rndm}{M}
\safemath{\rndn}{N}
\safemath{\rndo}{O}
\safemath{\rndp}{P}
\safemath{\rndq}{Q}
\safemath{\rndr}{R}
\safemath{\rnds}{S}
\safemath{\rndt}{T}
\safemath{\rndu}{U}
\safemath{\rndv}{V}
\safemath{\rndw}{W}
\safemath{\rndx}{X}
\safemath{\rndy}{Y}
\safemath{\rndz}{Z}

\safemath{\rveca}{\bimA}
\safemath{\rvecb}{\bimB}
\safemath{\rvecc}{\bimC}
\safemath{\rvecd}{\bimD}
\safemath{\rvece}{\bimE}
\safemath{\rvecf}{\bimF}
\safemath{\rvecg}{\bimG}
\safemath{\rvech}{\bimH}
\safemath{\rveci}{\bimI}
\safemath{\rvecj}{\bimJ}
\safemath{\rveck}{\bimK}
\safemath{\rvecl}{\bimL}
\safemath{\rvecm}{\bimM}
\safemath{\rvecn}{\bimN}
\safemath{\rveco}{\bomO}
\safemath{\rvecp}{\bimP}
\safemath{\rvecq}{\bimQ}
\safemath{\rvecr}{\bimR}
\safemath{\rvecs}{\bimS}
\safemath{\rvect}{\bimT}
\safemath{\rvecu}{\bimU}
\safemath{\rvecv}{\bimV}
\safemath{\rvecw}{\bimW}
\safemath{\rvecx}{\bimX}
\safemath{\rvecy}{\bimY}
\safemath{\rvecz}{\bimZ}

\safemath{\rvecxi}{\bmxi}
\safemath{\rveclambda}{\bmlambda}
\safemath{\rvecmu}{\bmmu}
\safemath{\rvectheta}{\bmtheta}
\safemath{\rvecphi}{\bmphi}

\safemath{\rmatA}{\bimA}
\safemath{\rmatB}{\bimB}
\safemath{\rmatC}{\bimC}
\safemath{\rmatD}{\bimD}
\safemath{\rmatE}{\bimE}
\safemath{\rmatF}{\bimF}
\safemath{\rmatG}{\bimG}
\safemath{\rmatH}{\bimH}
\safemath{\rmatI}{\bimI}
\safemath{\rmatJ}{\bimJ}
\safemath{\rmatK}{\bimK}
\safemath{\rmatL}{\bimL}
\safemath{\rmatM}{\bimM}
\safemath{\rmatN}{\bimN}
\safemath{\rmatO}{\bimO}
\safemath{\rmatP}{\bimP}
\safemath{\rmatQ}{\bimQ}
\safemath{\rmatR}{\bimR}
\safemath{\rmatS}{\bimS}
\safemath{\rmatT}{\bimT}
\safemath{\rmatU}{\bimU}
\safemath{\rmatV}{\bimV}
\safemath{\rmatW}{\bimW}
\safemath{\rmatX}{\bimX}
\safemath{\rmatY}{\bimY}
\safemath{\rmatZ}{\bimZ}

\safemath{\rmatDelta}{\bimDelta}
\safemath{\rmatLambda}{\bimLambda}
\safemath{\rmatPhi}{\bimPhi}
\safemath{\rmatSigma}{\bimSigma}
\safemath{\rmatOmega}{\bimOmega}
\safemath{\rmatTheta}{\bimTheta}

%% file: standard-macros.tex
\usepackage{amssymb}
\usepackage{amsfonts}
\usepackage{mathrsfs}
\usepackage{xspace}
\usepackage{bm}
\usepackage{fancyref}
\usepackage{textcomp}

\usepackage{multirow}
\usepackage{stmaryrd}

\newenvironment{textbmatrix}{	\setlength{\arraycolsep}{2.5pt}%
								\big[\begin{matrix}}{\end{matrix}\big]%
								\raisebox{0.08ex}{\vphantom{M}}}

\def\be{\begin{equation}}
\def\ee{\end{equation}}
\def\een{\nonumber \end{equation}}
\def\mat{\begin{bmatrix}}
\def\emat{\end{bmatrix}}
\def\btm{\begin{textbmatrix}}
\def\etm{\end{textbmatrix}}

\def\ba#1\ea{\begin{align}#1\end{align}}
\def\bas#1\eas{\begin{align*}#1\end{align*}}
\def\bs#1\es{\begin{split}#1\end{split}} 
\def\bg#1\eg{\begin{gather}#1\end{gather}}
\def\bml#1\eml{\begin{multline}#1\end{multline}}
\def\bi#1\ei{\begin{itemize}#1\end{itemize}}

\newcommand{\lefto}{\mathopen{}\left}

\DeclareMathOperator*{\argmin}{arg\;min}		

\newcommand{\vecnorm}[1]{\lefto\lVert#1\right\rVert}		

\safemath{\dirac}{\delta}					
\safemath{\krond}{\dirac}					

\safemath{\upto}{\uparrow}
\safemath{\downto}{\downarrow}
\safemath{\iu}{j}							
\safemath{\ev}{\lambda}						
\safemath{\hilseqspace}{l^{2}}				
\newcommand{\banachfunspace}[1]{\setL^{#1}}	
\safemath{\hilfunspace}{\banachfunspace{2}}	

\safemath{\SNR}{\textsf{SNR}} 				
\safemath{\PAR}{\textsf{PAR}} 				
\safemath{\No}{N_0}							
\safemath{\Es}{E_s}							
\safemath{\Eb}{E_b}							
\safemath{\EbNo}{\frac{\Eb}{\No}}
\safemath{\EsNo}{\frac{\Es}{\No}}

\DeclareMathOperator{\CHop}{\ensuremath{\opH}} 
\safemath{\tvir}{\rndh_{\CHop}}				
\safemath{\tvtf}{\rndl_{\CHop}}				
\safemath{\spf}{\rnds_{\CHop}}				
\safemath{\bff}{H_{\CHop}}					

\safemath{\ircf}{r_{h}}						
\safemath{\tftvcf}{r_{s}}					
\safemath{\tfcf}{r_{l}}						
\safemath{\bfcf}{r_{H}}						

\safemath{\tcorr}{c_h}						
\safemath{\scf}{c_{s}}						
\safemath{\tfcorr}{c_{l}}					
\safemath{\fcorr}{c_{H}}						

\safemath{\mi}{I}							
\safemath{\capacity}{C}						

\safemath{\normal}{\mathcal{N}}			
\safemath{\jpg}{\mathcal{CN}}			
\safemath{\mchain}{\leftrightarrow}		

\safemath{\dB}{\,\mathrm{dB}}
\safemath{\dBm}{\,\mathrm{dBm}}
\safemath{\Hz}{\,\mathrm{Hz}}
\safemath{\kHz}{\,\mathrm{kHz}}
\safemath{\MHz}{\,\mathrm{MHz}}
\safemath{\GHz}{\,\mathrm{GHz}}
\safemath{\s}{\,\mathrm{s}}
\safemath{\ms}{\,\mathrm{ms}}
\safemath{\mus}{\,\mathrm{\text{\textmu}s}}
\safemath{\ns}{\,\mathrm{ns}}
\safemath{\ps}{\,\mathrm{ps}}
\safemath{\meter}{\,\mathrm{m}}
\safemath{\mm}{\,\mathrm{mm}}
\safemath{\cm}{\,\mathrm{cm}}
\safemath{\m}{\,\mathrm{m}}
\safemath{\W}{\,\mathrm{W}}
\safemath{\mW}{\, \mathrm{mW}}
\safemath{\J}{\,\mathrm{J}}
\safemath{\K}{\,\mathrm{K}}
\safemath{\bit}{\,\mathrm{bit}}
\safemath{\nat}{\,\mathrm{nat}}

\safemath{\define}{\triangleq}			

\safemath{\equivalent}{\sim}
\safemath{\distas}{\sim}					
\safemath{\sdiff}{\Delta}				

\safemath{\reals}{\mathbb{R}}
\safemath{\positivereals}{\reals_{+}}
\safemath{\integers}{\mathbb{Z}}
\safemath{\posint}{\integers_{+}}
\safemath{\naturals}{\mathbb{N}}
\safemath{\posnaturals}{\naturals_{+}}
\safemath{\complexset}{\mathbb{C}}
\safemath{\rationals}{\mathbb{Q}}

\newcommand*{\fancyrefapplabelprefix}{app}		
\newcommand*{\fancyrefthmlabelprefix}{thm}		
\newcommand*{\fancyreflemlabelprefix}{lem}		
\newcommand*{\fancyrefcorlabelprefix}{cor}		
\newcommand*{\fancyrefdeflabelprefix}{def}		
\newcommand*{\fancyrefproplabelprefix}{prop}	
\newcommand*{\fancyrefobslabelprefix}{obs}		
\newcommand*{\fancyrefalglabelprefix}{alg}		
\newcommand*{\fancyrefremlabelprefix}{rem}		
\newcommand*{\fancyrefasmlabelprefix}{asm}	    

\frefformat{vario}{\fancyrefseclabelprefix}{Section~#1}
\frefformat{vario}{\fancyrefthmlabelprefix}{Theorem~#1}
\frefformat{vario}{\fancyreflemlabelprefix}{Lemma~#1}
\frefformat{vario}{\fancyrefcorlabelprefix}{Corollary~#1}
\frefformat{vario}{\fancyrefdeflabelprefix}{Definition~#1}
\frefformat{vario}{\fancyrefobslabelprefix}{Observation~#1}
\frefformat{vario}{\fancyrefasmlabelprefix}{Assumption~#1}
\frefformat{vario}{\fancyreffiglabelprefix}{Fig.~#1}
\frefformat{vario}{\fancyrefapplabelprefix}{Appendix~#1} 
\frefformat{vario}{\fancyrefproplabelprefix}{Proposition~#1}
\frefformat{vario}{\fancyrefalglabelprefix}{Algorithm~#1}
\frefformat{vario}{\fancyrefremlabelprefix}{Remark~#1}
\frefformat{vario}{\fancyrefeqlabelprefix}{(#1)}

%% file: defs.tex
\safemath{\dictab}{[\,\dicta\,\,\dictb\,]}

\safemath{\ysig}{\bmy}
\safemath{\ysighat}{\hat{\ysig}}
\safemath{\ysigdim}{M}
\safemath{\xsig}{\bmx}
\safemath{\xsigdim}{N}
\safemath{\nx}{n_x}
\safemath{\zsig}{\bmz}
\safemath{\zsigdim}{\ysigdim}
\safemath{\rsig}{\bmr}
\safemath{\Adict}{\bA}
\safemath{\Adicttilde}{\widetilde{\Adict}}
\safemath{\Adictdim}{\outputdim\times\xsigdim}
\safemath{\avec}{\bma}
\safemath{\avectilde}{\tilde{\avec}}
\safemath{\Bdict}{\bB}
\safemath{\Bdicttilde}{\widetilde{\Bdict}}
\safemath{\Cdict}{\bC}
\safemath{\cvec}{\bmc}
\safemath{\Ddict}{\bD}
\safemath{\Ddictdim}{\ysigdim\times\xsigdim}
\safemath{\dvec}{\bmd}
\safemath{\Ddicttilde}{\widetilde{\bD}}
\safemath{\Bonb}{\bB}
\safemath{\bvec}{\bmb}
\safemath{\Bonbdim}{\ysigdim\times\ysigdim}
\safemath{\noise}{\bmn}
\safemath{\noisedim}{\ysigim}
\safemath{\err}{\bme}
\safemath{\errdim}{\ysigdim}
\safemath{\errset}{\setE}
\safemath{\nerr}{n_e}
\safemath{\delop}{\bP_\errset}
\safemath{\delopc}{\bP_{{\errset}^c}}

\safemath{\cplxi}{\imath}
\safemath{\cplxj}{\jmath}

\safemath{\dict}{\matD}
\safemath{\inputdim}{N}		
\safemath{\outputdim}{M}		
\safemath{\sparsity}{S}	
\safemath{\inputdimA}{{N_a}}	
\safemath{\inputdimB}{{N_b}}	
\safemath{\elemA}{{n_a}}	
\safemath{\elemB}{{n_b}}	
\safemath{\resA}{\matR_a}	
\safemath{\resB}{\matR_b}	
\safemath{\subD}{\matS} 
\safemath{\subA}{\matS_a} 
\safemath{\subB}{\matS_b} 
\safemath{\dicta}{\matA} 	
\safemath{\dictb}{\matB} 	
\safemath{\hollowS}{H}
\safemath{\hollowA}{H_a}
\safemath{\hollowB}{H_b}
\safemath{\cross}{Z}
\safemath{\coh}{\mu_d}			
\safemath{\coha}{\mu_a}			
\safemath{\cohb}{\mu_b}			
\safemath{\mubs}{\nu}	
\safemath{\cohm}{\mu_m} 
\safemath{\dictset}{\setD}	
\safemath{\dictsetp}{\dictset(\coh,\coha,\cohb)}	
\safemath{\dictsetgen}{\dictset_\text{gen}}
\safemath{\dictsetgenp}{\dictsetgen(\coh)}
\safemath{\dictsetonb}{\dictset_\text{onb}}
\safemath{\dictsetonbp}{\dictsetonb(\coh)}

\safemath{\leftside}{U}
\safemath{\rightsideA}{R_a}
\safemath{\rightsideB}{R_b}

\safemath{\indexS}{\setI_S} 

\safemath{\na}{n_a}			
\safemath{\nb}{n_b}			
\safemath{\coeffa}{p_i}	
\safemath{\coeffb}{q_j}	
\safemath{\seta}{\setP}		
\safemath{\setb}{\setQ}     
\safemath{\setw}{\setW}	
\safemath{\setz}{\setZ}	
\safemath{\cola}{\veca}		
\safemath{\colb}{\vecb}		
\safemath{\cold}{\vecd}		
\safemath{\inputvec}{\vecx} 	
\safemath{\error}{\vece}	
\safemath{\noiseout}{\vecz} 	
\safemath{\inputvecel}{x}
\safemath{\inputveca}{\vecx_a}
\safemath{\inputvecb}{\vecx_b}
\safemath{\outputvec}{\vecy}	
\safemath{\lambdamin}{\lambda_{\mathrm{min}}}

\safemath{\elltwo}{\ell_2}
\safemath{\ellone}{\ell_1}
\safemath{\ellzero}{\ell_0}
\safemath{\ellinf}{\ell_\infty}
\safemath{\ellinftilde}{\ell_{\widetilde\infty}}
\safemath{\licard}{Z(\coh,\coha,\cohb)}
\safemath{\xsol}{\hat{x}}
\safemath{\xbord}{x_b}		
\safemath{\xstat}{x_s}		
\safemath{\xstatLone}{\tilde{x}_s}
\safemath{\order}{\mathcal{O}} 
\safemath{\scales}{\Theta} 
\safemath{\ones}{\mathbf{1}} 
\safemath{\zeroes}{\mathbf{0}} 
\safemath{\thlone}{\kappa(\coh,\cohb)} 
\safemath{\constoneA}{\delta} 
\safemath{\constoneB}{\epsilon} 
\safemath{\nlarge}{L}				   
\safemath{\sumlarge}{S_\nlarge}
\safemath{\maxlarger}{P_\nlarge}	   
\safemath{\Pzero}{\textrm{P0}}	
\safemath{\Pone}{\textrm{P1}}
\safemath{\vecfir}{\vecw}			 
\safemath{\vecsec}{\vecz}
\safemath{\elvecfir}{w}              
\safemath{\elvecsec}{z}				 
\safemath{\nlargefir}{n}
\safemath{\normout}{\gamma}
\safemath{\auxfun}{h}
\safemath{\supp}{\textrm{supp}}

\safemath{\indexa}{\ell}
\safemath{\indexb}{r}
\safemath{\indexc}{i}
\safemath{\indexd}{j}

\safemath{\project}{P}